\documentclass[aps,pra,twocolumn,showpacs,preprintnumbers,amsmath,amssymb,superscriptaddress]{revtex4}

\pdfoutput=1

\usepackage{graphicx}
\usepackage[latin1]{inputenc}
\usepackage{dcolumn}
\usepackage{bm}
\usepackage{mciteplus}

\newcommand{\ket}[1]{\left|#1\right\rangle}
\newcommand{\braket}[2]{\langle #1 \mid #2 \rangle}

\newtheorem{theorem}{Theorem}

\newtheorem{lemma}{Lemma}

\DeclareMathOperator{\Tr}{Tr}

\begin{document}

\title{Security of quantum key distribution with arbitrary individual imperfections}

\author{Øystein Marøy}
\email{oystein.maroy@iet.ntnu.no}
\affiliation{Department of Electronics and Telecommunications, Norwegian University of Science and
Technology, NO-7491 Trondheim, Norway}
\affiliation{University Graduate Center, NO-2027 Kjeller, Norway}

\author{Lars Lydersen}
\affiliation{Department of Electronics and Telecommunications, Norwegian University of Science and
Technology, NO-7491
Trondheim, Norway}
\affiliation{University Graduate Center, NO-2027 Kjeller, Norway}

\author{Johannes Skaar}
\affiliation{Department of Electronics and Telecommunications, Norwegian University of Science and
Technology, NO-7491
Trondheim, Norway}
\affiliation{University Graduate Center, NO-2027 Kjeller, Norway}

\date{\today}

\begin{abstract}
We consider the security of the Bennett-Brassard 1984 (BB84) protocol for quantum key distribution, with arbitrary individual imperfections simultaneously in the source and detectors. We provide the secure key generation rate, and show that three parameters must be bounded to ensure security; the basis dependence of the source, a detector blinding parameter, and a detector leakage parameter. The system may otherwise be completely uncharacterized and contain large losses. 
\end{abstract}


\maketitle

\section{Introduction}
Quantum Key Distribution (QKD) is a method for distributing a secure key to two communicating parties, Alice and Bob. The most common QKD protocol, BB84 \cite{bennett1984}, has been proved secure by a number of approaches, some of which include different kinds of imperfections in the equipment \cite{mayers1996,inamori2001,koashi2003,gottesman2004,fung2009,lydersen2008}. The ultimate goal of QKD security analysis is to take all kinds of imperfections into account, at least those that cannot be eliminated completely by a suitable design of the setup. So far, most of the available security proofs for BB84 consider imperfections at the source or detector separately. An exception is the work by Gottesman et. al. \cite{gottesman2004}, which treats the security in the presence of source flaws and a squashing detector with certain, limited imperfections. Also of interest is the paper by Hayashi \cite{Hayashi2007}, which combines finite length key analysis with photon number imperfections at the source. Proving security for a realistic system with arbitrary imperfections, simultaneously in the source, channel, and detectors, has so far been an open problem.

A particularly suitable approach for practical QKD is to limit the assumptions about the equipment. By considering entanglement-based protocols with detectors in both ends of the system \cite{ekert1991}, one can prove security in a rather general setting \cite{barrett2005}, assuming collective attacks and individual imperfections \cite{acin2007}. While these protocols and security proofs are promising, they do not necessarily provide security for realistic devices. All realistic systems have large losses due to the channel and limited detector efficiencies. An eavesdropper Eve may use imperfect detection efficiencies to effectively control Bob's basis choice \cite{makarov2006,makarov2007}. Using this detection-loophole, she may perform the identical measurement as Bob to obtain a perfect copy of the key \footnote{For any protocol, Bob's basis choice (or more generally, measurement setting) must be random and come from a trusted random number generator; otherwise Eve could perform the same measurement as Bob to obtain a perfect copy of his result.}. 

In this work we will prove security for BB84 with any combination of individual imperfections as well as channel losses. By individual imperfections we mean that the operation of the devices for a particular signal is independent of earlier signals. To obtain such generality, we describe the actual physics in the protocol, rather than using e.g. squashing models with ``tagging''. Thus, the detectors are described as a basis-dependent quantum operation on the actual state space, in front of a three-outcome measurement (``0'', ``1'', and ``vacuum''). Describing the detector in this way also enables an elegant solution to the problem of combining errors in the detectors and errors in the source.

To get around the detection loophole, we anticipate that at least two parameters must be known or bounded about the system; one for the source and one for the detectors. Our proof is formulated with two such parameters; the basis dependence of the source and a detector blinding parameter. In addition to these parameters, we include a third parameter quantifying leakage from Bob's detectors. Once these parameters are bounded, the system may contain bit and basis leakage from Alice, multimode behavior, basis-dependent misalignments, losses, nonlinearities, basis-dependent threshold detectors with detector efficiency mismatch and information leakage, dark counts, etc. In that sense, our proof offers the generality of the entanglement-based scenarios \cite{acin2007}, applies to realistic scenarios with loss, and provides universal composable security against the most general attacks.

\section{Protocol}
Consider the following BB84-like protocol, the actual protocol. Alice chooses basis $a=Z$ or $a=X$ randomly according to some probability distribution and prepares the state $|\chi_a\rangle$, where
\begin{subequations} \label{pure states}
\begin{align}
|\chi_Z\rangle&=\sqrt{p_Z}|0\rangle|\beta_0\rangle+\sqrt{1-p_Z}|1\rangle|\beta_1\rangle,\\
|\chi_X\rangle&=\sqrt{p_X}|+\rangle|\beta_+\rangle+\sqrt{1-p_X}|-\rangle|\beta_-\rangle. 
\end{align}
\end{subequations}
Here $p_Z$ and $p_X$ are probabilities, $|0\rangle$, $|1\rangle$ are some orthonormal qubit basis states, and $|\pm\rangle=(|0\rangle\pm|1\rangle)/\sqrt 2$. Alice measures the qubit in the $a$-basis (this measurement can be delayed to the end of the protocol). She repeats the procedure to obtain a large number of ``$\beta$ states'', which are sent via Eve to Bob. These $\beta$ states include any system that is correlated to Alice's system and to which Eve has access. Note that Eve is free to send anything to Bob, including parts of $\beta$ and/or any state of her own choice. Depending on Alice's source the four different $\beta$-states will differ in photon number statistics, polarization, wavelength, etc. Any leakage in non-photonic side channels will also be included in these states. With no loss of generality, the $\beta$-states are assumed pure; if they were mixed,
we could simply purify them, sending the auxiliary, purifying system to Eve.

For each state received by Bob, he chooses a ``basis'' variable $b$ according to some probability distribution and conducts measurements $M_b$. The measurements $M_b$ have three outcomes, ``0'', ``1'', and ``vacuum''. When he obtains ``0'' or ``1'' he publicly acknowledges receipt. After transmission, Alice and Bob broadcast $a$ and $b$. When $b=X$ they openly compare their measurement results to estimate the fraction $q_X$ of nonvacuum events at Bob when $a=X$, the corresponding error rate $\delta_X$, and the fraction $q_{\text{ph}}$ of nonvacuum events when $a=Z$. After this estimation only the $n$ states for which $a=b=Z$ are kept. Discarding all events where Bob detected ``vacuum'', Alice and Bob each end up with $nq_Z$ bits. Alice's bits are the raw key.

We will now summarize Koashi's generic framework for security proofs \cite{koashi2009, *koashi2005arxiv, koashi2006}. Imagine a virtual experiment where Alice measures her final $nq_Z$ qubits (corresponding to the raw key) in the $X$-basis instead of $Z$-basis. In this virtual experiment, instead of measuring $M_Z$, Bob now tries to predict the outcome of Alice's measurement. To do this, he may do whatever is permitted by quantum mechanics, as long as he does not alter the information given to Eve. Let $H_{\text{virt} X}(A|B=\mu)$ denote the entropy of Alice's result, given measurement result $\mu$ in Bob's prediction. Let $H_{\text{virt} X}(A|B=\mu)\leq H$ for some constant $H$. Since the uncertainty after Bob's prediction is less than $H$, the entropic uncertainty relation \cite{maassen1988} suggests that anyone (including Eve) cannot predict the outcome of a $Z$-basis measurement by Alice with less entropy than $nq_Z-H$. This indicates that Alice can extract $nq_Z-H$ bits of secret key. The quantity $H$ is to be found from the estimated parameters $q_X$, $\delta_X$, and $q_{\text{ph}}$ \footnote{The $Z$-basis error rate $\delta_Z$ is not needed to ensure that Alice's key is secret; thus there is no need to invoke the classicalization argument \cite{lo1999} regarding statistics of measurements involved in the simultaneous estimation of $\delta_X$ and $\delta_Z$.}. The detailed proof \cite{koashi2009} of the fact that Alice can extract $nq_Z-H$ bits of secret key is based upon the universal, composable security definition, and considers the actual privacy amplification protocol by universal hashing. 

To ensure that Bob has the identical key, we note that it does not matter to Eve what Bob does (as long as he gives the same receipt acknowledgment information); he can as well measure $M_Z$. Then Bob obtains the identical raw key from his measurement result and $nq_Zh(\delta_Z)$ extra bits of error correction information from Alice, consuming $nq_Zh(\delta_Z)$ of previous established secure key. Here $h(\cdot)$ is the binary Shannon entropy function, and the error rate $\delta_Z$ can be estimated by sacrificing a subset of the raw key (whose size we can neglect in the asymptotic limit $n\to\infty$). We therefore obtain the asymptotic net secure key generation rate 
\begin{equation}
\label{koashigeneralrate}
R_Z \geq 1-H/nq_Z-h(\delta_Z).
\end{equation}

\section{Individual imperfections in the detectors}
We first consider the situation where Alice's source is perfect ($\ket{\chi_X}=\ket{\chi_Z}$) and Bob's detectors can be subject to any kind of individual imperfections. With the understanding that Bob chooses his bit randomly for coincidence counts \cite{inamori2001,gottesman2004}, his detectors can be modeled by a basis-dependent quantum operation ($\mathcal E_Z$ and $\mathcal E_X$) in front of a measurement with three possible outcomes: ``0'', ``1'', and ``vacuum''. Note that there is no need to require a squash model \cite{gottesman2004,beaudry2008,tsurumaru2008} in our proof as Bob's basis selector is included into the basis-dependent quantum operation.

In addition to the optical modes, there may also be other relevant degrees of freedom in the detector. For example, dark counts are caused by physical processes internally in the detector. Thus we consider an extended state space consisting of the Fock space of all optical modes in addition to the state space associated with ``electronic'' degrees of freedom inside the detectors. Pessimistically, we let Eve control all degrees of freedom.

\begin{figure}
\includegraphics[width=8.6cm]{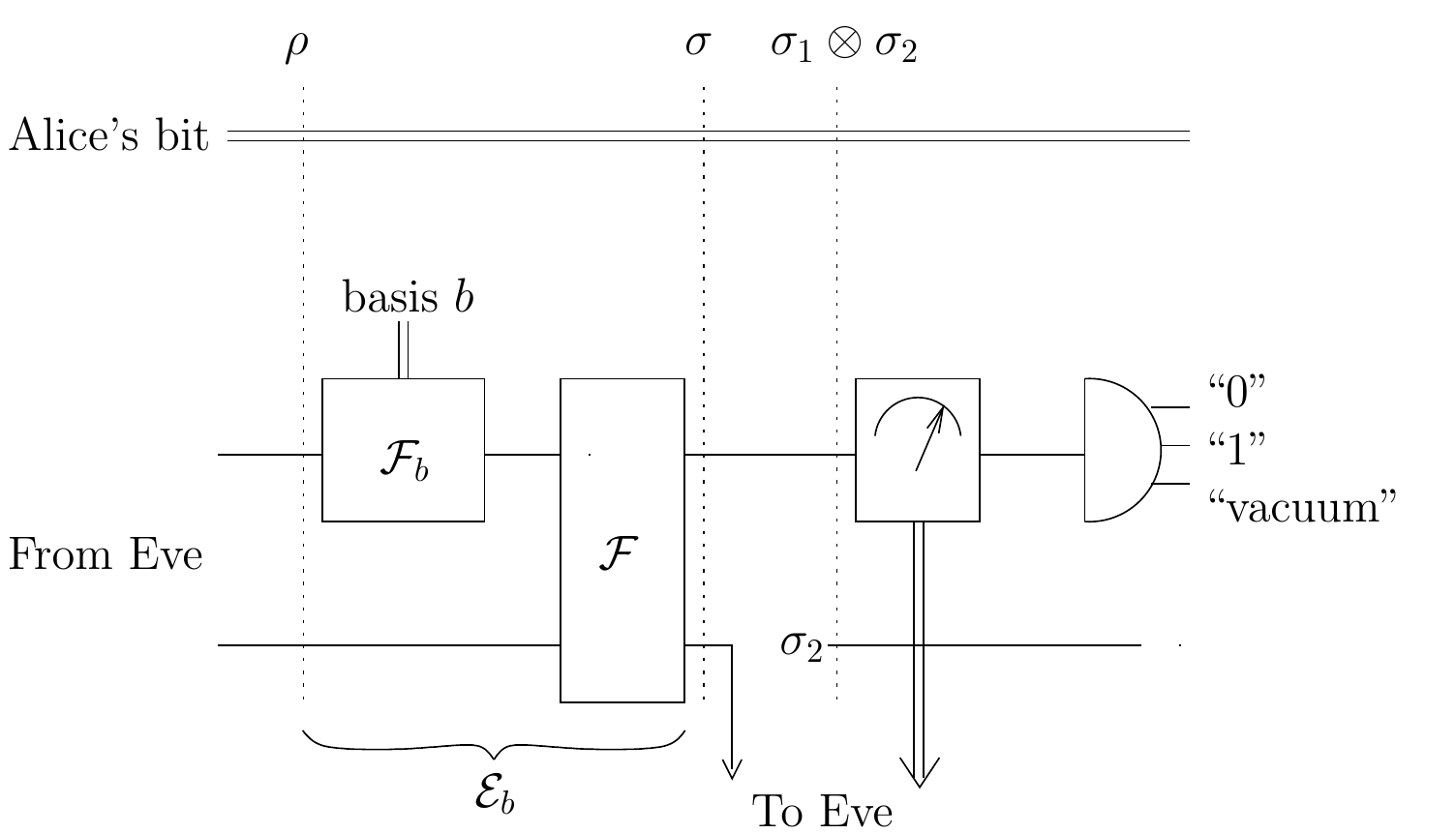}
\caption{\label{fig:detectormodel} Bob's detectors consist of a basis-dependent quantum operation ($\mathcal E_Z=\mathcal F\circ\mathcal F_Z$ and $\mathcal E_X=\mathcal F\circ\mathcal F_X$) in front of a three-outcome measurement. The fact that Eve gets arrival information from Bob is included through a dedicated vacuum measurement preceding Bob's three-outcome measurement. On the input side of $\mathcal F$, the lower line contains the electronic modes of the detector, while on the output side of $\mathcal F$, the lower line indicates the part of the Hilbert space leaked to Eve. Alice's classical bit, indicated in the upper part of the figure, is included in the state $\sigma$.}
\end{figure}

The quantum operations $\mathcal E_Z$ and $\mathcal E_X$ are decomposed as follows: First there is a basis-dependent quantum operation ($\mathcal F_Z$ and $\mathcal F_X$) acting on the Fock space associated with all optical modes. This operation contains Bob's basis selector. The operations $\mathcal F_Z$ and $\mathcal F_X$ are assumed passive in the sense that if vacuum is incident to all modes, there will also be vacuum at the output. Then there is another quantum operation $\mathcal F$ describing interaction between the photonic state and internal degrees of freedom in the detectors, see Fig. \ref{fig:detectormodel}. The quantum operation $\mathcal F$ may be active in the sense that even though vacuum is incident to all optical modes, there may be nonvacuum detections. When the optical modes contain the vacuum state, we can (pessimistically) assume that Eve has full control over Bob's detectors through $\mathcal F$; in other words, she controls the dark counts directly with the ``electronic'' modes. The quantum operation $\mathcal F$ is assumed to be independent of Bob's basis choice. This assumption is natural as Bob's basis choice does not influence internal degrees of freedom in the detector. In other words, when Eve emits the vacuum in all optical modes, Bob's basis choice will not affect the detection statistics.

To achieve a completely general detector model, we should not only let Eve control the detectors; in addition we must let information return to Eve. Consider the case where Bob has chosen the $Z$-basis. In the most general case the information leakage is quantum, i.e., a part of the total Hilbert space is given directly to Eve. Replacing this part of the Hilbert space by some standard state $\sigma_2$, we can quantify the leakage $\epsilon_Z$ by the trace distance $D(\cdot,\cdot)$ as follows:
\begin{equation}\label{epsilonleakage}
\epsilon_Z=\min_{\sigma_2} \max_{\rho} D(\sigma,\sigma_1\otimes\sigma_2).
\end{equation}
Here $\rho$ is any state at Bob's input (including Alice's part of the
system, see Fig. \ref{fig:detectormodel}), $\sigma$ is the state of
Alice and Bob before leakage, and $\sigma_1=\Tr_2(\sigma)$ is the state
of the remaining Hilbert space after leakage. Note that these density
operators refer to a single signal, not the entire block of $n$
signals. The parameter $\epsilon_Z$ measures the correlation between the leaked quantum state and the state of Alice and Bob, maximized over states sent by Eve. More precisely, $\epsilon_Z$ is the maximum probability that the actual state before leakage can be distinguished from the state where the leaked part is replaced by the standard state $\sigma_2$ \cite{nielsen2000}. Eq. \eqref{epsilonleakage} has another very useful physical interpretation: Choose a fixed $\sigma_2$, dependent on $\mathcal E_Z$, but independent of the state coming from Eve. For any $\sigma$, the probability of a measurement result of $\sigma_1\otimes\sigma_2$ deviates no more than $\epsilon_Z$ from the corresponding probability when measuring $\sigma$ \cite{nielsen2000}.

Although we now have a general detector model, we add one little feature. In the actual protocol, Eve gets to know whether a particular signal was detected. This can be included as an extra projective measurement with projectors $P$ and $I-P$, where $I-P$ is a projector onto the subspace corresponding to detection result ``vacuum'' in Bob's measurement. Clearly this addition does not disturb Bob's measurement statistics. The composed measurement consisting of $\mathcal E_Z$ followed by this projective measurement will be referred to as Eve's vacuum measurement. It can be described by some POVM elements $E$ and $I-E$, where $I-E$ corresponds to detection result ``vacuum'' at Bob. Including Eve's vacuum measurement separately, rather than absorbing it into the quantum leakage \eqref{epsilonleakage}, leads to a better rate. The reason is that the information from the vacuum measurement is classical and available to Bob, as opposed to general, leaked quantum information.

Having described the model, we now turn to the security analysis. As before, Alice extracts the key in the $Z$-basis. In Koashi's security proof, Bob wants to predict the outcome of a virtual $X$-basis measurement by Alice. In this virtual prediction there is only one important restriction: Bob is not allowed to alter the information going to Eve. Thus Eve's vacuum measurement must be retained.

\begin{figure}
\includegraphics[width=8.8cm]{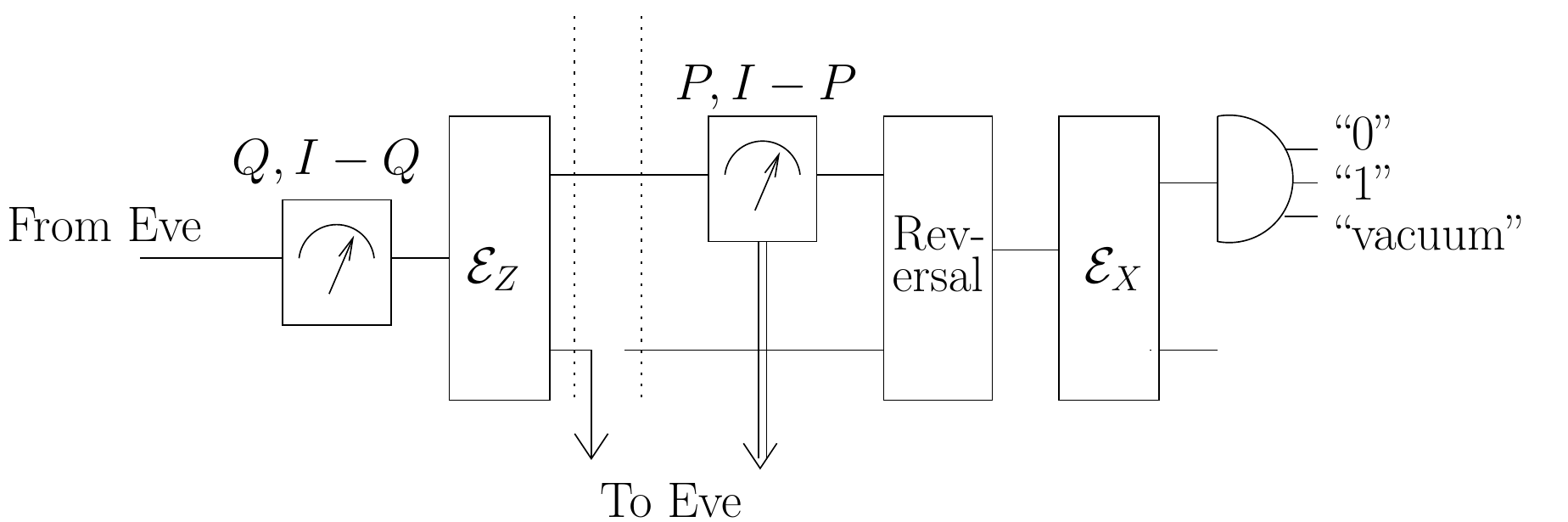}
\caption{\label{fig:virtualXsetup} Bob's setup for virtual $X$-basis prediction. The optical and electronic modes are denoted by a single line in this figure.}
\end{figure}

The setup used by Bob to perform the virtual $X$-basis prediction is depicted in Fig. \ref{fig:virtualXsetup}. The state from Eve is incident to a first vacuum measurement, Bob's vacuum measurement, a projective measurement with certain projectors $Q$ and $I-Q$, corresponding to results ``nonvacuum'' and ``vacuum'', respectively. Then it goes through the quantum operation $\mathcal E_Z$, and leaks partially back to Eve. The remaining part is measured by Eve's vacuum measurement, and sent through a reversal operation. The goal of the reversal operation is to reverse the effect of the vacuum measurement, so that the combined operation consisting of Eve's vacuum measurement and the reversal operation is identity, with a certain probability. Finally, the quantum operation $\mathcal E_X$ and Bob's three-outcome measurement are applied.

To analyze Bob's virtual prediction, we note the following observations. The quantum operation $\mathcal E_Z$ can be viewed as a unitary operation on an extended state space. Moreover, since Bob's reversal operation does not have to be realizable in practice (only in principle) we may assume that Bob has access to any extra degrees of freedom used to ``unitarize'' $\mathcal E_Z$. He does not have access to the quantum state leaked to Eve; however, the leakage disturbs the probabilities of Bob's prediction by no more than $\epsilon_Z$. Therefore, for the moment we can ignore the leakage, taking it into account in the final expression for the key rate.

To proceed, we need the following results.
\begin{lemma}\label{reversal} (Koashi and Ueda \cite{koashi1999}).
Let $E$, acting on a Hilbert space $\mathcal H$, be a POVM element associated with some measurement $M$. If any state in some subspace $\mathcal Q\subseteq\mathcal H$ is measured with $M$, the measured state can be reversed to the original state, with maximum joint probability of outcome $E$ and successful reversal $\inf_{|\Phi\rangle\in\mathcal Q, \langle\Phi|\Phi\rangle=1}\langle\Phi|E|\Phi\rangle$. It is possible to know when the reversal is successful or not.
\end{lemma}

\begin{lemma}\label{vacuummeasurement}
The output of a quantum operation $\mathcal E_b$ is measured with projectors $P_0$, $P_1$, and $I-P_0-P_1$, corresponding to detection results ``0'', ``1'', and ``vacuum'', respectively, or alternatively, with $P\equiv P_0+P_1$ and $I-P$. Let $I-Q$ be a projector onto an input subspace of $\mathcal E_b$ that leads to detection result ``vacuum'' with certainty. The measurement statistics are not changed by the presence of a projective measurement $\{Q,I-Q\}$ before $\mathcal E_b$.
\end{lemma}
Proof:
Lemma \ref{vacuummeasurement} is not as trivial as it may appear at first sight since states in the support of $Q$ may also lead to detection result ``vacuum''. Thus the measurement before $\mathcal E_b$ gives extra information. Nevertheless, the quantum operation $\mathcal E_b$ can be viewed as a unitary transformation on an extended Hilbert space, with a standard state as auxiliary input. Clearly, it does not matter if we measure the extra degrees of freedom at the output. This measurement can be constructed so that the total output measurement distinguishes between input states in the support of $Q$ or $I-Q$. Then, an input measurement $\{Q,I-Q\}$ is redundant. 

More precisely, the unitary operator can be chosen such that the projective measurement at the output is implemented as a measurement of a single qutrit in the computational basis. Thus it transforms
\begin{subequations}
\begin{align}
&|0_1\rangle|0\rangle_\text{aux}\to|v\rangle|\psi_1\rangle, \\
&|0_2\rangle|0\rangle_\text{aux}\to|v\rangle|\psi_2\rangle, 
\end{align}
\end{subequations}
and 
\begin{subequations}
\begin{align}
&|1_1\rangle|0\rangle_\text{aux}\to|v\rangle|\phi_1^v\rangle+|0\rangle|\phi_1^0\rangle+|1\rangle|\phi_1^1\rangle,\\ 
&|1_2\rangle|0\rangle_\text{aux}\to|v\rangle|\phi_2^v\rangle+|0\rangle|\phi_2^0\rangle+|1\rangle|\phi_2^1\rangle,
\end{align}
\end{subequations}
etc. Here $|0_i\rangle$ and $|1_i\rangle$ are bases for the support of $I-Q$ and $Q$, respectively, $|0\rangle_\text{aux}$ is the auxiliary standard state, and $|0\rangle\langle 0|=P_0$, $|1\rangle\langle 1|=P_1$, and $|v\rangle\langle v|=I-P_0-P_1$. The $\psi$- and $\phi$-vectors are (not necessarily normalized) states of the remaining part of the output state space. Since $\langle 1_i|0_j\rangle=0$, we have $\langle \phi_i^v|\psi_j\rangle=0$ for any $i,j$. Thus, by a measurement of the $\psi$ or $\phi$ part of the output state space in addition to the qutrit, we can distinguish between the $|0_i\rangle$ states and $|1_i\rangle$ states.$\,\square$

We define the projector $I-Q$ so as to project onto vacuum in all photonic modes, and onto the biggest subspace of the ``electronic'' modes that gives detection result ``vacuum'' in Eve's vacuum measurement. The orthogonal subspace, which is the support of $Q$, is denoted $\mathcal Q$. Lemma \ref{vacuummeasurement} ensures that Bob's vacuum measurement does not change the statistics of Eve's vacuum measurement. When Eve's vacuum measurement gives result ``vacuum'', or the reversal operation is not successful, the reversal operation is assumed to output a state in the support of $I-Q$. Thus in these cases the output of Bob's virtual prediction is ``vacuum'' with certainty.

If the outcome of Bob's vacuum measurement is ``vacuum'', the outcome of Eve's vacuum measurement is ``vacuum'', and the reversal operation is successful with certainty. Suppose the outcome of Bob's vacuum measurement is ``nonvacuum''. According to Lemma \ref{reversal}, the maximum joint probability of result $E$ in Eve's vacuum measurement and successful reversal is $\eta_Z=\inf_{|\Phi\rangle\in\mathcal Q, \langle\Phi|\Phi\rangle=1}\langle\Phi|E|\Phi\rangle$. When result $E$ and the reversal is successful (and Bob knows when it is), the statistics of Bob's measurement compared to Alice's virtual $X$-basis measurement will be identical to that of Alice's and Bob's ordinary parameter estimation in the $X$-basis, except for any disturbance by Bob's vacuum measurement. According to Lemma \ref{vacuummeasurement} such disturbance does not exist. The number of detection events $E$ in Eve's vacuum measurement is $nq_Z$; of these $nq_X\eta_Z$ is successfully reversed \emph{and} detected as ``0'' or ``1'' in Bob's virtual prediction. 
Thus we obtain $H\leq (nq_Z-nq_X\eta_Z) + nq_X\eta_Z h(\delta_X)$, which gives us the rate
\begin{equation} \label{rateeta}
R_Z \geq \eta_Z q_X/q_Z(1-h(\delta_X))-h(\delta_Z).
\end{equation}

The parameter $\eta_Z=\inf_{|\Phi\rangle\in\mathcal Q,
\langle\Phi|\Phi\rangle=1}\langle\Phi|E|\Phi\rangle$ is the minimum probability that a state in $\mathcal Q$ gives result $E$ by Eve. This parameter has a clear physical interpretation. When vacuum is incident to the optical modes, recall that with no loss of generality we may assume that Eve has full control of the detectors through the ``electronic'' modes. Then there are no losses of her excitation in the ``electronic'' modes through the quantum operation $\mathcal F$. Thus, we identify $\eta_Z$ as the minimum probability that a nonvacuum photonic state is detected by Bob. In other words, $1-\eta_Z$ is the maximum probability that a nonvacuum photonic state is absorbed in the detectors and detected as vacuum in the actual setup (Fig. \ref{fig:detectormodel}).

So far we have ignored the effect of any quantum leakage from the detectors. Parameterizing the leakage by \eqref{epsilonleakage}, $\epsilon_Z$ quantifies the maximum deviation of any measurement probabilities. In the absence of leakage, the probabilities of correct and incorrect predictions are $q_X\eta_Z(1-\delta_X)$ and $q_X\eta_Z\delta_X$, respectively, while the probability of vacuum result is $1-q_X\eta_Z$. When there is leakage, in the worst case these probabilities are changed to $q_X\eta_Z(1-\delta_X)-\epsilon_Z$, $q_X\eta_Z\delta_X+\epsilon_Z-\xi$, and $1-q_X\eta_Z+\xi$, respectively. Here $\xi$ is an unknown parameter satisfying $0\leq\xi\leq \epsilon_Z$. Of the $nq_Z$ nonvacuum results in Eve's vacuum measurement, there are $n(q_X\eta_Z-\xi)$ nonvacuum results in Bob's virtual prediction. This leads to
\begin{align}
H & \leq nq_Z-n(q_X\eta_Z-\xi) \nonumber\\
  & + n(q_X\eta_Z-\xi) h\left(\frac{q_X\eta_Z\delta_X+\epsilon_Z-\xi}{q_X\eta_Z-\xi}\right)
\nonumber\\
  & \leq nq_Z-nq_X\eta_Z + nq_X\eta_Z h\left(\delta_X+\frac{\epsilon_Z}{q_X\eta_Z}\right).
\label{Hleakage}
\end{align}
The last inequality in \eqref{Hleakage} can be found after some algebra using the facts that $h(u)-h(u-\Delta) \geq h'(u)\Delta$ for $\Delta\geq 0$ and $u\leq 1/2$, and $h'(u)(1-u)\geq 1$ for $u\leq0.277$. Here we have set $u=\delta_X+\frac{\epsilon_Z}{q_X\eta_Z}$. 

This gives the rate
\begin{equation} \label{rateetaleakage}
R_Z \geq \eta_Z \frac{q_X}{q_Z}\left[1-h\left(\delta_X+\frac{\epsilon_Z}{q_X\eta_Z}\right)\right]-h(\delta_Z),
\end{equation}
for $\delta_X+\frac{\epsilon_Z}{q_X\eta_Z}\leq 0.277$. An expression for the rate, also valid for $0.277\leq\delta_X+\frac{\epsilon_Z}{q_X\eta_Z}\leq 0.5$, can be derived straightforwardly; however, this regime is only relevant for very small $\delta_Z$, and large $\delta_X$ and/or $\frac{\epsilon_Z}{q_X\eta_Z}$.

\section{Individual imperfections in the entire system}
From the previous section we note that when the reversal operation is successful (and Bob knows when it is), the measurement statistics in the prediction becomes identical to the statistics if Bob measured in the $X$-basis. This makes it possible to consider simultaneous imperfections at the source and detector. We may then consider the case where Alice creates a general state $\rho_a$ depending on the basis choice $a$. The basis dependence of the source is characterized by the fidelity $F(\rho_Z, \rho_X)\equiv\Tr(\sqrt{\rho_Z}\rho_X\sqrt{\rho_Z})^{\frac{1}{2}}$. We let this dependence be bounded by a parameter $\Delta$ defined by $F\geq1-2\Delta$. By Uhlmann's theorem there exist purifications, $\ket{\chi_a}$ of $\rho_a$, such that $\braket{\chi_Z}{\chi_X}=1-2\Delta$. We note that $\ket{\chi_a}$ can be expressed as in Eq. \eqref{pure states}.

Again, we first ignore the detector leakage, taking it into account in the final expression for the rate. Since Bob wants to predict Alice's virtual X-basis measurement on $\ket{\chi_Z}$, the parameters $\delta_X$ and $q_X$ in \eqref{rateeta} must be replaced with $\delta_{\text{ph}}$ and $q_{\text{ph}}$ respectively. Here $\delta_{\text{ph}}$ is the error rate when Alice measures her part of $\ket{\chi_Z}$ in the $X$-basis and Bob measures his part using $M_X$. 

In BB84 such a measurement is not actually performed, but $\delta_{\text{ph}}$ can be bounded from the measured error and transmission rates. We expand the statistical argument from \cite{koashi2009} to include ''vacuum'' as a possible measurement result. Assume that for the systems used in the random sampling Alice chooses her basis by measuring a quantum coin in the $Z$-basis. Then these systems can be described by state $\ket{\Psi}=(\ket{\chi_Z}\ket{0}+\ket{\chi_X}\ket{1})/\sqrt{2}$, with the last system being that of the quantum coin. 

We then consider the situations where Alice and Bob both conduct $X$-basis measurements. For each measurement a variable $t$ is assigned the value $t=0$ if their results are the same, $t=1$ if there is an error, and $t=2$ if Bob gets no result. Alice then measures her quantum coin in the $Z$-basis, getting the result $c$. We obtain the following conditional probabilities.
\begin{subequations}\label{cond prob}
\begin{align}
&p(t=0|c=1)=q_X(1-\delta_X)\\ &p(t=0|c=0)=q_\text{ph}(1-\delta_\text{ph}) \\
&p(t=1|c=1)=q_X\delta_X\\ &p(t=1|c=0)=q_\text{ph}\delta_\text{ph} \\
&p(t=2|c=1)=1-q_X\\ &p(t=2|c=0)=1-q_\text{ph}.
\end{align}
\end{subequations}
Assuming that the systems used to estimate error and transmission rates are randomly chosen, the probabilities given $c=0$ are also valid for the systems used to extract the raw key.
 
Now assume that for some states Alice measure the coin in the X-basis getting measurement result $\bar{c}$. Note that
\begin{equation} \label{coin measurement}
\sum_j p(t=j)p(\bar{c}=1|t=j)=\Delta.
\end{equation}
Using \eqref{cond prob}, \eqref{coin measurement} and the bound \cite{tamaki2003}
\[(1-2p(\bar{c}=1|t=j))^2+(1-2p(c=0|t=j))^2\leq 1\]
we find
\begin{align}\label{deltaph}
&1-2\Delta\leq \sum_j\sqrt{p(t=j|a=Z)p(t=j|a=X)}\\\nonumber
&=\sqrt{q_X(1-\delta_X)q_\text{ph}(1-\delta_\text{ph})}+\sqrt{q_X\delta_X q_\text{ph}\delta_\text{ph}}\\\nonumber
&+\sqrt{(1-q_X)(1-q_\text{ph})}.
\end{align}
$\delta_\text{ph}$ can now be taken to be the maximal value for which the inequality is obeyed. 

Similarly to the analysis in the previous section, we can include detector leakage by modifying the detection probabilities. As in \eqref{rateetaleakage}, the leakage is accounted for by adding a term proportional to the leakage parameter $\epsilon_Z$, 
\begin{equation} \label{deltaphleakage}
\tilde{\delta}_\text{ph}\leq\delta_\text{ph}+\frac{\epsilon_Z}{q_\text{ph}\eta_Z}.
\end{equation}
We have arrived at our main result.
\begin{theorem}
In BB84 the basis-dependence of Alice's source is bounded by $F(\rho_X,\rho_Z)\geq1-2\Delta$. Bob's detectors are modeled by a passive, basis-dependent quantum operation ($\mathcal F_Z$ and $\mathcal F_X$) acting on the multimode photonic state, followed by a basis-independent quantum operation ($\mathcal F$) describing interaction with internal degrees of freedom in the physical detector, followed by a measurement with three outcomes ``0'', ``1'', and ``vacuum''. Suppose Eve controls the photonic modes and the internal degrees of freedom in the detectors, and that a quantum state leaks back to Eve from the detectors. Then the asymptotic secure key generation rate for key extraction in the $Z$-basis satisfies
\begin{equation}
\label{finalrate}
R_Z \geq \eta_Z q_{\textnormal{ph}}/q_Z\left[1-h(\tilde{\delta}_{\textnormal{ph}})\right]-h(\delta_Z),
\end{equation}
provided $\tilde{\delta}_{\textnormal{ph}}\leq 0.277$. Here $\delta_Z$ is the estimated error rate in the $Z$-basis, $\tilde{\delta}_{\textnormal{ph}}$ is given by \eqref{deltaph} and \eqref{deltaphleakage}, $1-\eta_Z$ is the maximum probability that a non-vacuum photonic state is detected as ``vacuum'', and $q_{\textnormal{ph}}/q_Z$ is the ratio between the detection rates for Bobs measurements $M_X$ and $M_Z$ given that Alice sends in the Z-basis. 
\end{theorem}
The rate \eqref{finalrate} is valid for any kinds of individual imperfections and loss. The parameters $q_X$, $q_Z$, $q_\text{ph}$, $\delta_X$, and $\delta_Z$ are estimated directly in the protocol, while $\Delta$, $\eta_Z$, and $\epsilon_Z$ characterize the practical setup.

\section{Discussion of results}
In this discussion we assume that the quantum channel is symmetric with respect to loss i.e. $q_X=q_\text{ph}=q_Z\equiv q$. This will be approximately true for most setups. We also assume no information returned to Eve from the detectors, $\epsilon_Z=0$, anticipating that such errors could be avoided by modifying the setup.

\begin{figure} 
\includegraphics[width=8.6cm]{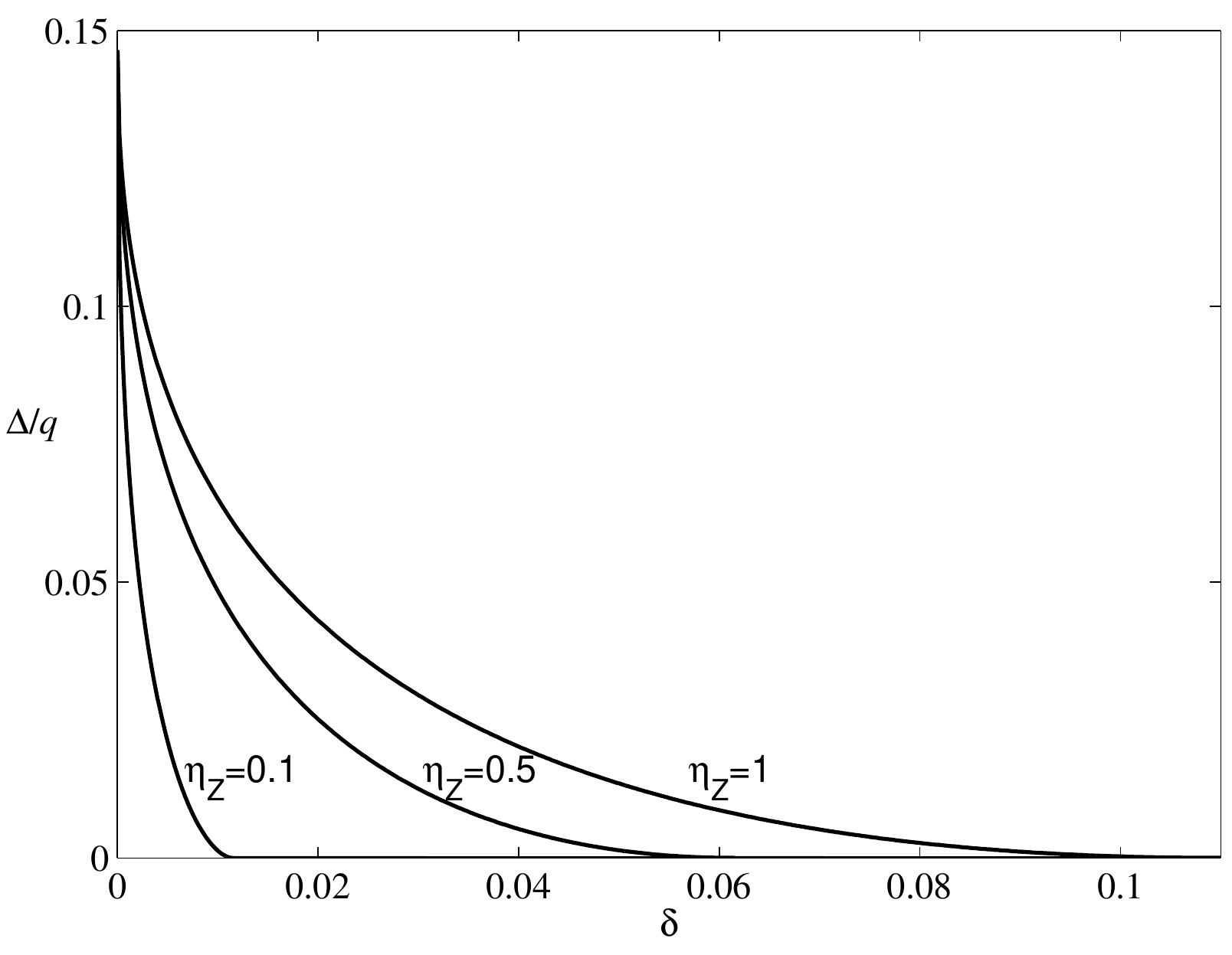}
\includegraphics[width=8.6cm]{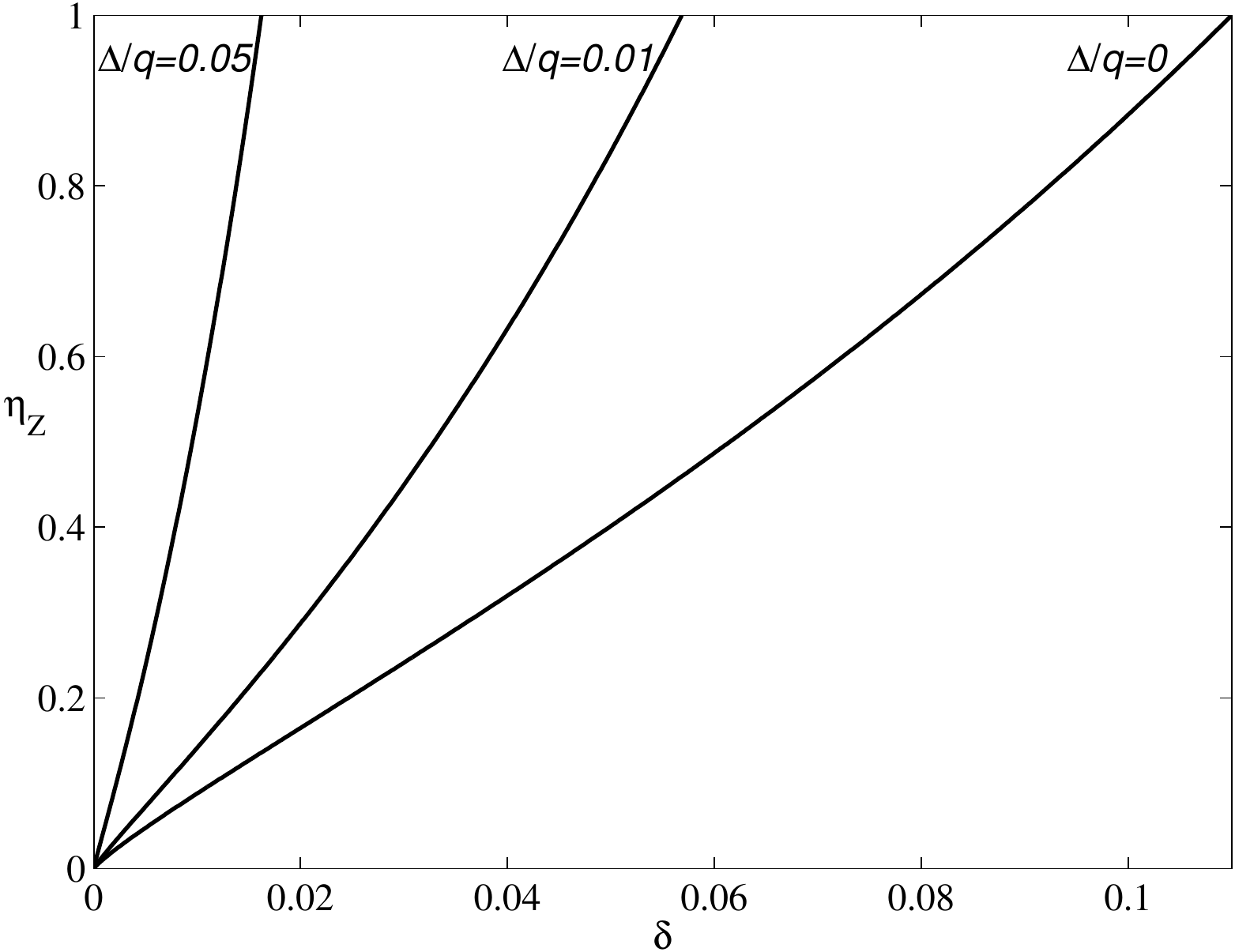}
\includegraphics[width=8.6cm]{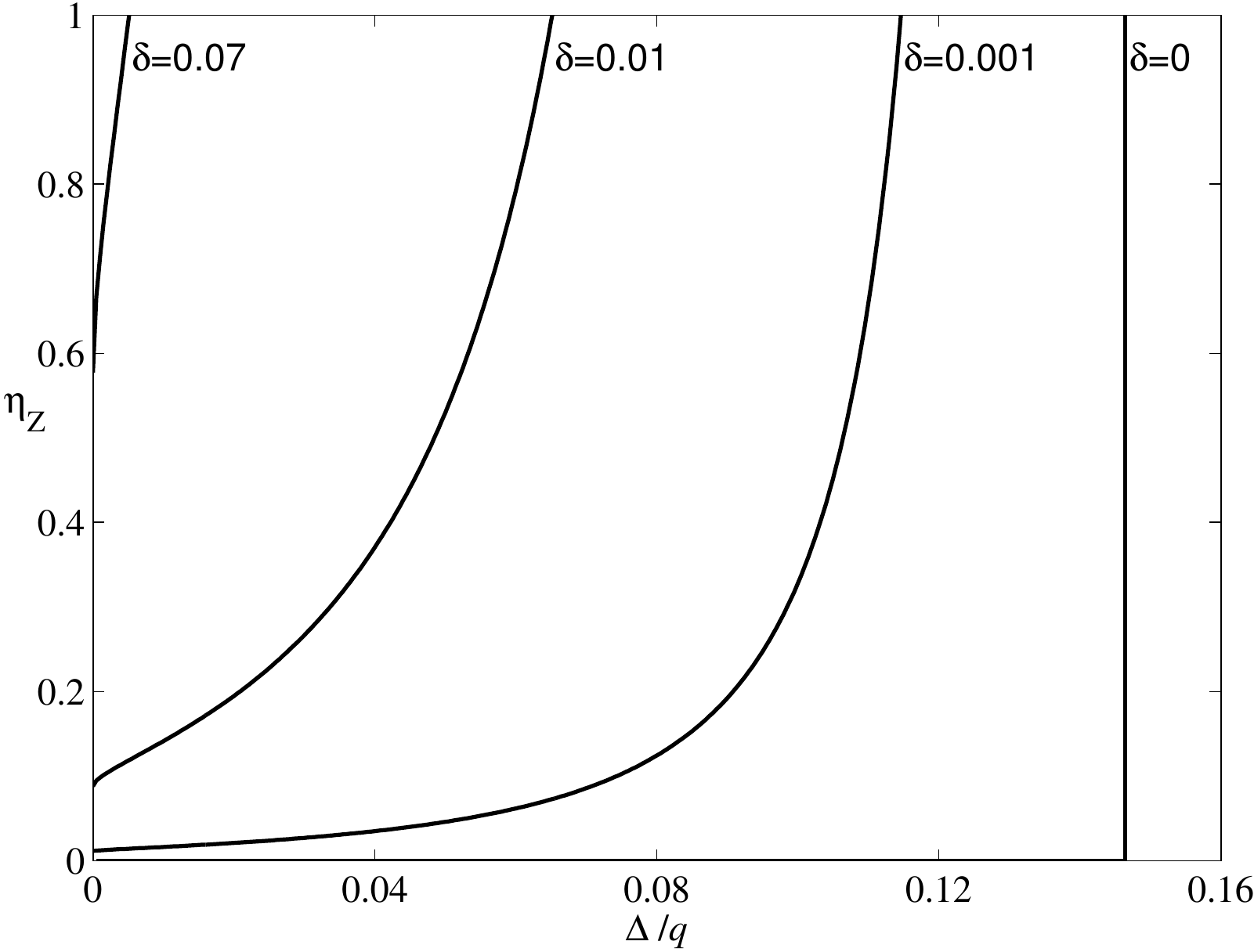}
\caption{\label{fig:Keygain} Plots showing the security bounds $R_Z=0$ for different values of $\eta_Z$, $\delta$, and $\frac{\Delta}{q}$. The security bound is found by setting $R_Z=0$ in \eqref{finalrate}. Positive key gain is possible for parameter values to the left of the curves. We have assumed $\epsilon_Z=0,$ $\delta_X=\delta_Z=\delta$ and $q_X=q_\text{ph}=q_Z=q$.}
\end{figure}

In this case \eqref{deltaph} reduces to
\begin{equation}
\frac{2\Delta}{q}\geq1-\sqrt{(1-\delta_X)(1-\delta_\text{ph})}-\sqrt{\delta_X\delta_\text{ph}} 
\end{equation}
and the estimated worst possible error rate is: 
\begin{equation}\begin{split}
\delta_\text{ph}=&\min\Biggl\{\frac{1}{2}\;,\;\delta_X+8\frac{\Delta}{q}\Biggl((1-\frac{\Delta}{q})(1-2\delta_X)
\\&+\sqrt{\frac{\Delta}{q}(1-\frac{\Delta}{q})\delta_X(1-\delta_X)}\Biggr)\Biggr\}
\end{split}\end{equation}
We see that errors in the source are more critical when the transmission is low. This is due to Eves control of the channel, which let her pass to Bob only the systems where her operation has given her the most information for the least disturbance. If the source is perfect, $\Delta=0$, loss in the channel does not affect the secret key rate. The upper limit on the source error for which key gain is possible is $\frac{\Delta}{q}\leq \frac{\sqrt{2}-1}{2\sqrt{2}}\approx0.146$. This is independent of the detector parameter $\eta_Z$, as long as it is nonzero, but demands error rates equal to zero. For larger error rates the limit depends heavily on $\eta_Z$, Fig. \ref{fig:Keygain}. 

Channel loss and imperfect sources only contributes to an increase in $\delta_{\text{ph}}$. A better estimate of $\delta_{\text{ph}}$ would increase the rate. This is related to the method of decoy states \cite{hwang2003,lo2005,wang2005a}, where Alice instead of producing $\rho_Z$, sometimes produces a decoy state with a different mean photon number. From the transmission and error rates for this state, Alice and Bob are able to derive a stricter bound on $\delta_{\text{ph}}$ effectively reducing $R_Z$'s dependence of channel loss. To generalize this method, using decoy states where other properties of the signal state are varied, might prove useful when operating with an imperfect source. However creating such states may require the detailed output statistics of the source, and might be experimentally difficult in general. 

Considering the special case of a perfect source, our rate is larger than the rate proved for restricted detector flaws in previous literature \cite{lydersen2008,fung2009}. Key gain is possible for $\eta_Z\leq\frac{h(\delta_Z)}{1-h(\delta_X)}$. Unlike previous results, our rate applies to all relevant, individual imperfections at the detectors; for example, mode coupling including misalignments and multiple reflections, nonlinearities, mode dependent losses and detector efficiency mismatch, and any basis dependence of those effects. Moreover it applies to threshold detectors with dark counts.

Note that the detector blinding parameter $\eta_Z$ is not supposed to contain the transmission efficiency of the channel. Generally one should factorize $\mathcal E_Z=\tilde{\mathcal E}_Z\circ\mathcal E$ and $\mathcal E_X=\tilde{\mathcal E}_X\circ\mathcal E$ to put as much as possible of the imperfections into the basis-independent operation $\mathcal E$. By absorbing $\mathcal E$ into Eve and treating $\tilde{\mathcal E}_Z$ and $\tilde{\mathcal E}_X$ as the new imperfections, $\eta_Z$ will be maximal. For example, for the case where reduced detector efficiencies can be described as beamsplitters in front of ideal detectors, and if there is no coupling between modes associated with different logical bits, $\eta_Z$ is the minimum ratio between the two detection efficiencies \cite{lydersen2008}. For detectors that cannot be modeled by beamsplitters in front of ideal detectors, our security proof clearly shows the danger associated with the possibility of detector blinding \cite{makarov2007}: If the detection probability of a nonvacuum state is zero, our proof predicts zero key rate. For the case where the detectors can only be partially blinded, our proof can predict positive rate.

Returning to the general case, the rate is dependent on $\Delta$, $\eta_Z$, and $\epsilon_Z$, in addition to estimated parameters. For a specific QKD setup, $\Delta$ and $\epsilon_Z$ must be upper bounded, and $\eta_Z$ and must be lower bounded. How to deal with this in practice, is an interesting question for future research.

\section{Conclusion}
We have proved security for arbitrary, individual imperfections in a BB84 system. The detector model includes a basis dependent quantum operation, possibly with quantum leakage back to Eve, followed by a three-outcome measurement with outcomes ``0'', ``1'', and ``vacuum''. Such a general detector model can describe detector efficiency mismatch, nonlinear blindable behavior, response to multiple modes, mode coupling and multiple reflections, misalignments, back-reflection leakage, non-optical leakage, etc. By reversal of the receipt acknowledgment measurement on Bob's side (Eve's vacuum measurement), we show how to treat the general case with a lossy channel and general, individual imperfections at the source, combined with the flawed detector.


\begin{thebibliography}{26}
\expandafter\ifx\csname natexlab\endcsname\relax\def\natexlab#1{#1}\fi
\expandafter\ifx\csname bibnamefont\endcsname\relax
  \def\bibnamefont#1{#1}\fi
\expandafter\ifx\csname bibfnamefont\endcsname\relax
  \def\bibfnamefont#1{#1}\fi
\expandafter\ifx\csname citenamefont\endcsname\relax
  \def\citenamefont#1{#1}\fi
\expandafter\ifx\csname url\endcsname\relax
  \def\url#1{\texttt{#1}}\fi
\expandafter\ifx\csname urlprefix\endcsname\relax\def\urlprefix{URL }\fi
\providecommand{\bibinfo}[2]{#2}
\providecommand{\eprint}[2][]{\url{#2}}

\bibitem[{\citenamefont{Bennett and Brassard}(1984)}]{bennett1984}
\bibinfo{author}{\bibfnamefont{C.~H.} \bibnamefont{Bennett}} \bibnamefont{and}
  \bibinfo{author}{\bibfnamefont{G.}~\bibnamefont{Brassard}}, in
  \emph{\bibinfo{booktitle}{Proceedings of IEEE International Conference on
  Computers, Systems, and Signal Processing}} (\bibinfo{publisher}{IEEE Press,
  New York}, \bibinfo{address}{Bangalore, India}, \bibinfo{year}{1984}), pp.
  \bibinfo{pages}{175--179}.

\bibitem[{\citenamefont{Mayers}(1996)}]{mayers1996}
\bibinfo{author}{\bibfnamefont{D.}~\bibnamefont{Mayers}}, in
  \emph{\bibinfo{booktitle}{Proceedings of Crypto´96}}, edited by
  \bibinfo{editor}{\bibfnamefont{N.}~\bibnamefont{Koblitz}}
  (\bibinfo{publisher}{Springer, New York}, \bibinfo{year}{1996}), vol.
  \bibinfo{volume}{1109}, pp. \bibinfo{pages}{343--357}.

\bibitem[{\citenamefont{Inamori et~al.}(2001)\citenamefont{Inamori, Lütkenhaus,
  and Mayers}}]{inamori2001}
\bibinfo{author}{\bibfnamefont{H.}~\bibnamefont{Inamori}},
  \bibinfo{author}{\bibfnamefont{N.}~\bibnamefont{Lütkenhaus}},
  \bibnamefont{and} \bibinfo{author}{\bibfnamefont{D.}~\bibnamefont{Mayers}},
  \bibinfo{journal}{e-print quant-ph/0107017}  (\bibinfo{year}{2001}).

\bibitem[{\citenamefont{Koashi and Preskill}(2003)}]{koashi2003}
\bibinfo{author}{\bibfnamefont{M.}~\bibnamefont{Koashi}} \bibnamefont{and}
  \bibinfo{author}{\bibfnamefont{J.}~\bibnamefont{Preskill}},
  \bibinfo{journal}{Phys. Rev. Lett.} \textbf{\bibinfo{volume}{90}},
  \bibinfo{pages}{057902} (\bibinfo{year}{2003}).

\bibitem[{\citenamefont{Gottesman et~al.}(2004)\citenamefont{Gottesman, Lo,
  Lütkenhaus, and Preskill}}]{gottesman2004}
\bibinfo{author}{\bibfnamefont{D.}~\bibnamefont{Gottesman}},
  \bibinfo{author}{\bibfnamefont{H.-K.} \bibnamefont{Lo}},
  \bibinfo{author}{\bibfnamefont{N.}~\bibnamefont{Lütkenhaus}},
  \bibnamefont{and} \bibinfo{author}{\bibfnamefont{J.}~\bibnamefont{Preskill}},
  \bibinfo{journal}{Quantum Information \& Computation}
  \textbf{\bibinfo{volume}{4}}, \bibinfo{pages}{325} (\bibinfo{year}{2004}).

\bibitem[{\citenamefont{Fung et~al.}(2009)\citenamefont{Fung, Tamaki, Qi, Lo,
  and Ma}}]{fung2009}
\bibinfo{author}{\bibfnamefont{C.-H.~F.} \bibnamefont{Fung}},
  \bibinfo{author}{\bibfnamefont{K.}~\bibnamefont{Tamaki}},
  \bibinfo{author}{\bibfnamefont{B.}~\bibnamefont{Qi}},
  \bibinfo{author}{\bibfnamefont{H.-K.} \bibnamefont{Lo}}, \bibnamefont{and}
  \bibinfo{author}{\bibfnamefont{X.}~\bibnamefont{Ma}},
  \bibinfo{journal}{Quantum Information \& Computation}
  \textbf{\bibinfo{volume}{9}}, \bibinfo{pages}{131} (\bibinfo{year}{2009}).

\bibitem[{\citenamefont{Lydersen and Skaar}(2010)}]{lydersen2008}
\bibinfo{author}{\bibfnamefont{L.}~\bibnamefont{Lydersen}} \bibnamefont{and}
  \bibinfo{author}{\bibfnamefont{J.}~\bibnamefont{Skaar}},
  \bibinfo{journal}{Quant. Inf. Comp} \textbf{\bibinfo{volume}{10}},
  \bibinfo{pages}{0060} (\bibinfo{year}{2010}).

\bibitem[{\citenamefont{Hayashi}(2007)}]{Hayashi2007}
\bibinfo{author}{\bibfnamefont{M.}~\bibnamefont{Hayashi}},
  \bibinfo{journal}{Phys. Rev. A} \textbf{\bibinfo{volume}{76}},
  \bibinfo{pages}{012329} (\bibinfo{year}{2007}).

\bibitem[{\citenamefont{Ekert}(1991)}]{ekert1991}
\bibinfo{author}{\bibfnamefont{A.~K.} \bibnamefont{Ekert}},
  \bibinfo{journal}{Phys. Rev. Lett.} \textbf{\bibinfo{volume}{67}},
  \bibinfo{pages}{661} (\bibinfo{year}{1991}).

\bibitem[{\citenamefont{Barrett et~al.}(2005)\citenamefont{Barrett, Hardy, and
  Kent}}]{barrett2005}
\bibinfo{author}{\bibfnamefont{J.}~\bibnamefont{Barrett}},
  \bibinfo{author}{\bibfnamefont{L.}~\bibnamefont{Hardy}}, \bibnamefont{and}
  \bibinfo{author}{\bibfnamefont{A.}~\bibnamefont{Kent}},
  \bibinfo{journal}{Phys. Rev. Lett.} \textbf{\bibinfo{volume}{95}},
  \bibinfo{pages}{010503} (\bibinfo{year}{2005}).

\bibitem[{\citenamefont{Ac\'{\i}n et~al.}(2007)\citenamefont{Ac\'{\i}n,
  Brunner, Gisin, Massar, Pironio, and Scarani}}]{acin2007}
\bibinfo{author}{\bibfnamefont{A.}~\bibnamefont{Ac\'{\i}n}},
  \bibinfo{author}{\bibfnamefont{N.}~\bibnamefont{Brunner}},
  \bibinfo{author}{\bibfnamefont{N.}~\bibnamefont{Gisin}},
  \bibinfo{author}{\bibfnamefont{S.}~\bibnamefont{Massar}},
  \bibinfo{author}{\bibfnamefont{S.}~\bibnamefont{Pironio}}, \bibnamefont{and}
  \bibinfo{author}{\bibfnamefont{V.}~\bibnamefont{Scarani}},
  \bibinfo{journal}{Phys. Rev. Lett.} \textbf{\bibinfo{volume}{98}},
  \bibinfo{eid}{230501} (\bibinfo{year}{2007}).

\bibitem[{\citenamefont{Makarov et~al.}(2006)\citenamefont{Makarov, Anisimov,
  and Skaar}}]{makarov2006}
\bibinfo{author}{\bibfnamefont{V.}~\bibnamefont{Makarov}},
  \bibinfo{author}{\bibfnamefont{A.}~\bibnamefont{Anisimov}}, \bibnamefont{and}
  \bibinfo{author}{\bibfnamefont{J.}~\bibnamefont{Skaar}},
  \bibinfo{journal}{Physical Review A} \textbf{\bibinfo{volume}{74}},
  \bibinfo{pages}{022313} (\bibinfo{year}{2006}), \bibinfo{note}{ibid.
  \textbf{78}, 019905 (2008)}.

\bibitem[{\citenamefont{Makarov}(2009)}]{makarov2007}
\bibinfo{author}{\bibfnamefont{V.}~\bibnamefont{Makarov}},
  \bibinfo{journal}{New J. Phys.} \textbf{\bibinfo{volume}{11}},
  \bibinfo{pages}{065003} (\bibinfo{year}{2009}).

\bibitem[{\citenamefont{Koashi}(2009)}]{koashi2009}
\bibinfo{author}{\bibfnamefont{M.}~\bibnamefont{Koashi}}, \bibinfo{journal}{New
  Jornal of Physics} \textbf{\bibinfo{volume}{11}}, \bibinfo{pages}{045018}
  (\bibinfo{year}{2009}).
\bibinfo{author}{\bibfnamefont{M.}~\bibnamefont{Koashi}},
  \bibinfo{journal}{e-print quant-ph/0505108v1}  (\bibinfo{year}{2005}).

\bibitem[{\citenamefont{Koashi}(2006)}]{koashi2006}
\bibinfo{author}{\bibfnamefont{M.}~\bibnamefont{Koashi}},
  \bibinfo{journal}{e-print quant-ph/0609180}  (\bibinfo{year}{2006}).

\bibitem[{\citenamefont{Maassen and Uffink}(1988)}]{maassen1988}
\bibinfo{author}{\bibfnamefont{H.}~\bibnamefont{Maassen}} \bibnamefont{and}
  \bibinfo{author}{\bibfnamefont{J.~B.~M.} \bibnamefont{Uffink}},
  \bibinfo{journal}{Phys. Rev. Lett.} \textbf{\bibinfo{volume}{60}},
  \bibinfo{pages}{1103} (\bibinfo{year}{1988}).

\bibitem[{\citenamefont{Beaudry et~al.}(2008)\citenamefont{Beaudry, Moroder,
  and Lütkenhaus}}]{beaudry2008}
\bibinfo{author}{\bibfnamefont{N.~J.} \bibnamefont{Beaudry}},
  \bibinfo{author}{\bibfnamefont{T.}~\bibnamefont{Moroder}}, \bibnamefont{and}
  \bibinfo{author}{\bibfnamefont{N.}~\bibnamefont{Lütkenhaus}},
  \bibinfo{journal}{Phys. Rev. Lett.} \textbf{\bibinfo{volume}{101}},
  \bibinfo{pages}{093601} (\bibinfo{year}{2008}).

\bibitem[{\citenamefont{Tsurumaru and Tamaki}(2008)}]{tsurumaru2008}
\bibinfo{author}{\bibfnamefont{T.}~\bibnamefont{Tsurumaru}} \bibnamefont{and}
  \bibinfo{author}{\bibfnamefont{K.}~\bibnamefont{Tamaki}},
  \bibinfo{journal}{Phys. Rev. A} \textbf{\bibinfo{volume}{78}},
  \bibinfo{pages}{032302} (\bibinfo{year}{2008}).

\bibitem[{\citenamefont{Nielsen and Chuang}(2000)}]{nielsen2000}
\bibinfo{author}{\bibfnamefont{M.~A.} \bibnamefont{Nielsen}} \bibnamefont{and}
  \bibinfo{author}{\bibfnamefont{I.~L.} \bibnamefont{Chuang}},
  \emph{\bibinfo{title}{Quantum computation and quantum information}}
  (\bibinfo{publisher}{Cambridge University Press},
  \bibinfo{address}{Cambridge}, \bibinfo{year}{2000}).

\bibitem[{\citenamefont{Koashi and Ueda}(1999)}]{koashi1999}
\bibinfo{author}{\bibfnamefont{M.}~\bibnamefont{Koashi}} \bibnamefont{and}
  \bibinfo{author}{\bibfnamefont{M.}~\bibnamefont{Ueda}},
  \bibinfo{journal}{Phys. Rev. Lett.} \textbf{\bibinfo{volume}{82}},
  \bibinfo{pages}{2598} (\bibinfo{year}{1999}).

\bibitem[{\citenamefont{Tamaki et~al.}(2003)\citenamefont{Tamaki, Koashi, and
  Imoto}}]{tamaki2003}
\bibinfo{author}{\bibfnamefont{K.}~\bibnamefont{Tamaki}},
  \bibinfo{author}{\bibfnamefont{M.}~\bibnamefont{Koashi}}, \bibnamefont{and}
  \bibinfo{author}{\bibfnamefont{N.}~\bibnamefont{Imoto}},
  \bibinfo{journal}{Phys. Rev. Lett.} \textbf{\bibinfo{volume}{90}},
  \bibinfo{pages}{167904} (\bibinfo{year}{2003}).

\bibitem[{\citenamefont{Hwang}(2003)}]{hwang2003}
\bibinfo{author}{\bibfnamefont{W.~Y.} \bibnamefont{Hwang}},
  \bibinfo{journal}{Phys. Rev. Lett.} \textbf{\bibinfo{volume}{91}},
  \bibinfo{pages}{057901} (\bibinfo{year}{2003}).

\bibitem[{\citenamefont{Lo et~al.}(2005)\citenamefont{Lo, Ma, and
  Chen}}]{lo2005}
\bibinfo{author}{\bibfnamefont{H.-K.} \bibnamefont{Lo}},
  \bibinfo{author}{\bibfnamefont{X.~F.} \bibnamefont{Ma}}, \bibnamefont{and}
  \bibinfo{author}{\bibfnamefont{K.}~\bibnamefont{Chen}},
  \bibinfo{journal}{Phys. Rev. Lett.} \textbf{\bibinfo{volume}{94}},
  \bibinfo{pages}{230504} (\bibinfo{year}{2005}).

\bibitem[{\citenamefont{Wang}(2005)}]{wang2005a}
\bibinfo{author}{\bibfnamefont{X.-B.} \bibnamefont{Wang}},
  \bibinfo{journal}{Phys. Rev. Lett.} \textbf{\bibinfo{volume}{94}},
  \bibinfo{eid}{230503} (\bibinfo{year}{2005}).

\bibitem[{\citenamefont{Lo and Chau}(1999)}]{lo1999}
\bibinfo{author}{\bibfnamefont{H.-K.} \bibnamefont{Lo}} \bibnamefont{and}
  \bibinfo{author}{\bibfnamefont{H.~F.} \bibnamefont{Chau}},
  \bibinfo{journal}{Science} \textbf{\bibinfo{volume}{283}},
  \bibinfo{pages}{2050} (\bibinfo{year}{1999}).

\end{thebibliography}
\end{document}